\documentclass[10pt,aps,prd,twocolumn,showpacs,superscriptaddress,longbibliography,nofootinbib]{revtex4-2}
\usepackage{mathrsfs,amsmath,amsthm,latexsym,amssymb,amsfonts,epsfig,cancel,enumerate,graphicx,txfonts,diagbox}
\usepackage[utf8]{inputenc}  
\usepackage[T1]{fontenc}     
\usepackage{lmodern}         
\usepackage[percent]{overpic}
\usepackage{multirow}
\usepackage[T1]{fontenc}
\usepackage[utf8]{inputenc}
\usepackage[colorlinks=true,linkcolor=blue,citecolor=blue,urlcolor=blue]{hyperref}
\usepackage{orcidlink}

\usepackage{xcolor}

\usepackage{booktabs}
\usepackage{graphicx}
\definecolor{navy}{RGB}{0,0,150}
\usepackage{appendix}
\allowdisplaybreaks

\newcommand{\RGUI}{Department of Physics, The Assam Royal Global University, Guwahati-781035, Assam, India}
\newcommand{\UCCB}{Programa de P\'os-Gradua\c c\~ao em F\'{\i}sica \& Coordena\c c\~ao do Curso de F\'{\i}sica -- Bacharelado, Universidade Federal do Maranh\~{a}o, 65085-580 S\~{a}o Lu\'{\i}s, Maranh\~{a}o, Brazil}

\begin{document}
\baselineskip=14pt

\title{Hawking Temperature, Sparsity and Energy Emission Rate of Dark Matter Halo Regular Black Holes}

\author{Faizuddin Ahmed\orcidlink{0000-0003-2196-9622}}
\email{faizuddinahmed15@gmail.com}
\affiliation{\RGUI}
\author{Edilberto O. Silva\orcidlink{0000-0002-0297-5747}}
\email{edilberto.silva@ufma.br}
\affiliation{\UCCB}

\begin{abstract}
In this paper, we investigate the thermodynamic and radiative properties of a regular black hole sourced by a dark matter halo described by the Einasto density profile. The closed-form expressions for the Hawking temperature, specific heat capacity, sparsity parameter of Hawking flux, and the spectral energy emission rate were obtained. All these are examined as a function of the characteristic scale parameter $\alpha$ of the dark matter distribution and compared with the standard Schwarzschild results. We show that the presence of a dark matter halo suppresses both the Hawking temperature and energy emission rate
relative to the standard black hole result. Crucially, the specific heat capacity can be positive for a finite range of horizon radii, signalling a thermodynamically stable phase; the boundary of this stable region defines a Davies-type phase transition. The sparsity parameter is enhanced compared to the standard black hole, indicating that the dark matter environment renders the Hawking flux even more intermittent.\\

\end{abstract}

\maketitle

\section{Introduction}\label{sec:1}

The direct detection of gravitational waves-beginning with the landmark GW150914 event and followed by an expanding catalog from the LIGO-Virgo-KAGRA (LVK) network-has opened a new observational window onto the strong-field, highly dynamical regime of gravity \cite{Abbott2016GW,Abbott2025GWTC3GR}. In contrast to Solar System experiments, which have tested general relativity (GR) primarily in weak-field, quasi-static conditions \cite{Will2014GR}, compact-binary coalescences provide a unique laboratory for probing GR in regimes where the gravitational field is both nonlinear and rapidly evolving \cite{Yunes2025GW}. In addition, recent observations by the Event Horizon Telescope (EHT) of the supermassive black holes M87* \cite{EHTL1} and Sgr~A* \cite{EHTL12} have provided direct images of black hole shadows, offering compelling evidence for the validity of GR in the strong-gravity regime.

Black holes in realistic astrophysical settings are not expected to be completely isolated. Galactic dark matter halos can modify the spacetime geometry in the vicinity of compact objects in a nontrivial way. Due to the strong gravitational potential of black holes, dark matter can
be gravitationally captured and form a dense structure around the black holes
\cite{Gondolo1999DM,Eda2013DMGW,Bar2019ULDM,Mack2007PBH}, making the black hole neighborhood an ideal laboratory to study dark matter
indirectly. Dense dark matter environments can be classified according to
their density profiles, which typically correspond to different formation
mechanisms. Various approaches have been proposed to probe dark matter in
such scenarios
\cite{Akil2023DMProbe,Kadota2024SIDM,Aurrekoetxea2024WaveDM,Ding2025PBHDM}.
The presence of a dense dark matter environment surrounding a black hole
can significantly modify the gravitational-wave waveform during the
inspiral phase \cite{Eda2015Minispikes,Ghodla2025LISA,Chen2026DMGW}.
Precise measurements of gravitational-wave signals, therefore, offer a unique
opportunity to investigate dark matter properties in the strong-field
gravity regime. Dark matter with various density profiles has been
extensively investigated in the literature, including studies of geodesic
structure, thermodynamic properties, quasinormal modes, greybody factors,
and black-hole shadows by numerous researchers
(see, e.g.,
\cite{Konoplya2021GalacticCenters,AlBadawi2026,Ahmed2025,
AlBadawi2025Dehnen,AlBadawi2025QNM,AlBadawi2025JCAP}).

Among the various phenomenological models used to describe dark matter
halos, the Einasto density profile has received significant attention
\cite{Einasto1965,EinastoHaud1989}. Originally developed in the context
of galactic structure, this profile provides an accurate fit to dark
matter distributions obtained from numerical simulations of cosmic
structure formation and has become a standard tool in astrophysical
modelling
\cite{Springel2008,Navarro2010,Roszkowski2018,DuttonMaccio2014,
CatenaUllio2010,Navarro1997,Bertone2005,Hernquist1990}.
In this context, embedding black holes within spacetimes sourced by such
density profiles offers a natural framework to study how environmental
effects may influence black-hole observables and associated physical
properties.

In a recent work, Konoplya and Zhidenko \cite{Konoplya2026} constructed
exact black hole solutions that are free from curvature singularities and
are sourced by dark matter halos modeled through realistic galactic
density profiles, including Einasto and Dehnen-type distributions. Related
extensions have also considered charged black holes surrounded by galactic
halos in the presence of a cosmological constant
\cite{Konoplya2025ChargedHalo}. In this framework, the regularity of the
spacetime geometry is achieved by imposing the relation $P_r = -\rho$
between the radial pressure and the energy density, consistent with the
phenomenological flexibility of halo models. It was further demonstrated
that these regular black hole solutions remain stable under axial
perturbations. In a related study, the quasinormal modes of the regular
black hole sourced by the Einasto density profile was analyzed
\cite{Skvortsova2026}. The line element describing this regular black hole
solution is given by \cite{Konoplya2026,Skvortsova2026}
\begin{equation}
    ds^2 = -f(r)\,dt^2 + \frac{dr^2}{f(r)} + r^2\,d\Omega^2,
    \label{metric}
\end{equation}
where $d\Omega^2 = d\theta^2 + \sin^2\!\theta\,d\varphi^2$ and the lapse
function reads
\begin{align}
     f(r)= 1 - \frac{2 M}{r} +M\left(\frac{2}{r}+\frac{2}{\alpha}
     +\frac{r}{\alpha^2}\right) e^{-r/\alpha}.\label{function}
\end{align}
Here, the parameter $\alpha$ sets the characteristic radial scale of the
dark matter halo, and $M$ represents the black hole mass. Near the origin, the density tends to a constant value and the metric function behaves as
\begin{equation}
    f(r) \simeq 1-\Lambda_{\rm eff}\,r^2+\mathcal{O}(r^3),
    \qquad \Lambda_{\rm eff}>0,
\end{equation}
with an effective de~Sitter curvature scale determined by the dark matter
halo central density $\rho_0$, ensuring the geometry is regular at the
origin. In the limit $\alpha\to 0^+$, one recovers the Schwarzschild lapse $f(r) = 1 - 2M/r$ for any fixed $r>0$, since the exponential halo term vanishes outside the center.

Motivated by these developments, the present work investigates the
thermodynamic and radiative properties of such regular black holes,
including the Hawking temperature, specific heat capacity, and sparsity of
Hawking radiation and energy emission rate, with a systematic analysis
of the role played by $\alpha$.

In 1974, Hawking \cite{Hawking1975} demonstrated that a black hole behaves
approximately like a blackbody, emitting particles thermally at a
temperature proportional to its surface gravity. However, as measured by
an asymptotic observer, the Hawking emission spectrum is not an exact
Planckian distribution; this deviation arises from the curved spacetime
geometry between the event horizon and infinity, which allows only a
fraction of the emitted radiation to reach distant observers. An important
feature of Hawking radiation is its intrinsic sparsity
\cite{Gray2016,Hod2016PLB,Hod2015EPJC,Miao2017PLB,Schuster2018Thesis,Visser2017MGM}.
The emission process is highly intermittent: the typical time interval
between successive Hawking quanta is much larger than the characteristic
timescale associated with the emission of an individual quantum. As a
result, Hawking radiation can be described not as a continuous flux but
as a sequence of well-separated emission events. Understanding how a dark
matter environment modifies all of these properties is the primary
motivation of the present paper.

This paper is organized as follows. In Sec.~\ref{sec:2} we compute the
Hawking temperature and specific heat capacity. In Sec.~\ref{sec:3} we
analyze the sparsity of the Hawking flux. In Sec.~\ref{sec:4} we derive
the energy emission rate via the shadow radius. Concluding remarks are
presented in Sec.~\ref{sec:5}.\\

\section{Hawking Temperature and Specific Heat}\label{sec:2}

We investigate the thermodynamics of the considered black hole by deriving
the Hawking temperature and the specific heat capacity. Since the metric
function depends on the mass parameter $M$, and the horizon condition
relates $M$ to the horizon radius, special care is required when taking
derivatives along the horizon branch.

The surface gravity is given by \cite{Hawking1975,BardeenCarterHawking1973,Wald1994}
\begin{equation}
    \kappa = -\frac{1}{2}\lim_{r \to r_h}
    \frac{\partial_r g_{tt}}{\sqrt{-g_{tt}\,g_{rr}}}
    = \frac{1}{2}\left(\frac{\partial f}{\partial r}\right)_{M,\alpha}
    \bigg|_{r=r_h}.
    \label{aa1}
\end{equation}
Therefore, the Hawking temperature $T_H=\kappa/(2\pi)$ is
\begin{align}
T_H
&= \frac{1}{4\pi}
\left(\frac{\partial f}{\partial r}\right)_{M,\alpha}
\bigg|_{r=r_h}
= \frac{1}{4\pi} \Bigg[\frac{2M}{r_h^2}
    + M e^{-r_h/\alpha}\notag\\
&\times
    \left(
    -\frac{2}{r_h^2}
    + \frac{1}{\alpha^2}
    - \frac{1}{\alpha}
    \left(
    \frac{2}{r_h}
    +\frac{2}{\alpha}
    +\frac{r_h}{\alpha^2}
    \right)
    \right)\Bigg].
    \label{aa2}
\end{align}
Here the derivative in Eq.~\eqref{aa2} is a partial derivative with
respect to $r$, taken at fixed $M$ and fixed $\alpha$.

The horizon condition is defined by
\begin{equation}
    f(r_h,M,\alpha)=0.
\end{equation}
Solving this condition for the mass parameter gives the horizon-branch
mass
\begin{equation}
M_h(r_h,\alpha)=
\left[
\frac{2}{r_h}
-\left(
\frac{2}{r_h}
+\frac{2}{\alpha}
+\frac{r_h}{\alpha^2}
\right)e^{-r_h/\alpha}
\right]^{-1}.
\label{mass}
\end{equation}
Thus, on the horizon branch, the temperature must be understood as a
function of $r_h$ and $\alpha$ through
\begin{equation}
    T_H(r_h,\alpha)
    =
    \frac{1}{4\pi}\,
    \mathcal F(r_h,\alpha),
\end{equation}
where we define
\begin{equation}
    \mathcal F(r_h,\alpha)
    \equiv
    \left(\frac{\partial f}{\partial r}\right)_{M,\alpha}
    \bigg|_{r=r_h,\,M=M_h(r_h,\alpha)} .
    \label{F_def}
\end{equation}
Explicitly,
\begin{align}
\mathcal F(r_h,\alpha)
&=
\frac{2M_h}{r_h^2}
+ M_h e^{-r_h/\alpha}\notag\\
&\times 
\Bigg(
-\frac{2}{r_h^2}
+\frac{1}{\alpha^2}
-\frac{1}{\alpha}
\left(
\frac{2}{r_h}
+\frac{2}{\alpha}
+\frac{r_h}{\alpha^2}
\right)
\Bigg).
\label{F_explicit}
\end{align}

\begin{figure}[htbp]
\centering
\includegraphics[width=\columnwidth]{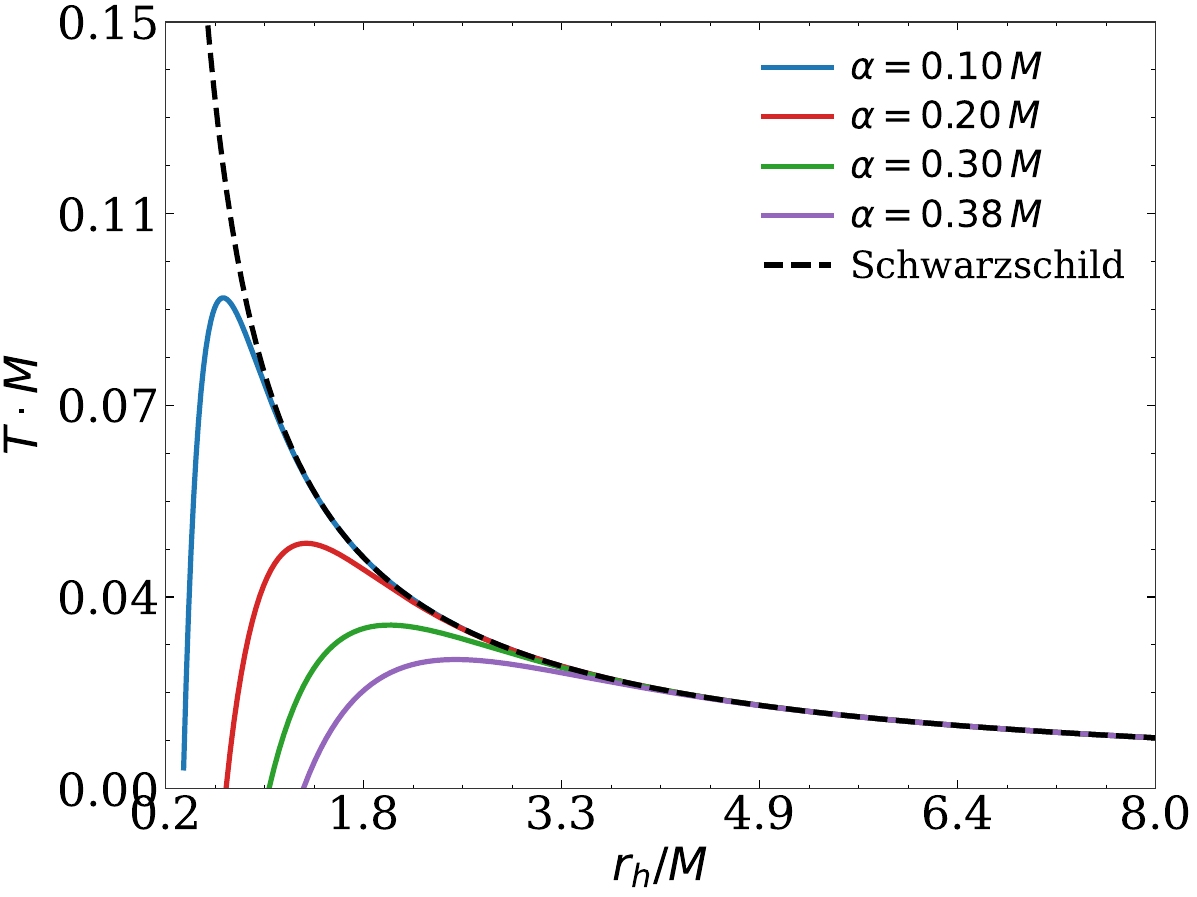}
\caption{Hawking temperature $T$ as a function of the horizon radius
$r_h$ (in units of $M$) for representative values
$\alpha/M=0.10,\,0.20,\,0.30,$ and $0.38$. The dashed black curve
corresponds to the Schwarzschild black hole,
$T_{\rm Sch}=1/(4\pi r_h)$. Larger $\alpha$ leads to a stronger
suppression of the temperature, while the $\alpha\to 0$ limit recovers
the Schwarzschild result.}
\label{fig:temperature}
\end{figure}

In the limit $\alpha\to 0^+$, the exponential halo term vanishes for
any fixed $r_h>0$, and one recovers the Schwarzschild result
$T_{\rm Sch}=1/(8\pi M)=1/(4\pi r_h)$. For finite $\alpha$, the dark
matter halo reduces the temperature below the Schwarzschild value for the
same horizon radius, with the suppression becoming more pronounced as
$\alpha$ increases, as illustrated in Fig.~\ref{fig:temperature}. Each
regular branch starts at a minimum allowed horizon radius, reaches a
maximum, and then approaches the Schwarzschild behavior for
$r_h\gg \alpha$.

We now compute the specific heat capacity. Adopting the Bekenstein--Hawking entropy \cite{Bekenstein1973,Hawking1975}
\begin{equation}
    S=\pi r_h^2,
\end{equation}
the specific heat must be evaluated along the horizon branch
$M=M_h(r_h,\alpha)$:
\begin{equation}
    C
    =
    T_H\frac{dS}{dT_H}
    =
    T_H
    \frac{dS/dr_h}{dT_H/dr_h}.
\end{equation}
Using
\begin{equation}
    \frac{dS}{dr_h}=2\pi r_h,
    \qquad
    T_H=\frac{\mathcal F}{4\pi},
\end{equation}
one obtains
\begin{equation}
    C
    =
    2\pi r_h\,
    \frac{\mathcal F(r_h,\alpha)}
    {\dot{\mathcal F}(r_h,\alpha)},
    \label{aa2a}
\end{equation}
where
\begin{equation}
    \dot{\mathcal F}(r_h,\alpha)
    \equiv
    \frac{d\mathcal F(r_h,\alpha)}{dr_h}
\end{equation}
denotes a total derivative along the horizon branch, not a partial
derivative at fixed mass.

To make this point explicit, differentiating the horizon condition
$f(r_h,M_h(r_h,\alpha),\alpha)=0$ with respect to $r_h$ gives
\begin{equation}
    \left(\frac{\partial f}{\partial r}\right)_{M,\alpha}
    +
    \left(\frac{\partial f}{\partial M}\right)_{r,\alpha}
    \frac{dM_h}{dr_h}
    =0.
\end{equation}
Since the metric function can be written as
\begin{equation}
    f(r,M,\alpha)=1+M A(r,\alpha),
\end{equation}
with
\begin{equation}
    A(r,\alpha)
    =
    -\frac{2}{r}
    +
    \left(
    \frac{2}{r}
    +\frac{2}{\alpha}
    +\frac{r}{\alpha^2}
    \right)e^{-r/\alpha},
\end{equation}
the horizon condition implies $A(r_h,\alpha)=-1/M_h$. Therefore,
\begin{equation}
    \frac{dM_h}{dr_h}
    =
    M_h\,\mathcal F(r_h,\alpha).
    \label{Mprime}
\end{equation}

Now, because $\mathcal F$ is evaluated on the horizon branch, its
derivative is
\begin{align}
    \dot{\mathcal F}(r_h,\alpha)
    &=
    \frac{d}{dr_h}
    \left[
    \left(\frac{\partial f}{\partial r}\right)_{M,\alpha}
    \right]_{M=M_h}
    \notag\\
    &=
    \left(\frac{\partial^2 f}{\partial r^2}\right)_{M,\alpha}
    +
    \left(\frac{\partial^2 f}{\partial M\,\partial r}\right)_{\alpha}
    \frac{dM_h}{dr_h}.
\end{align}
Since $f(r,M,\alpha)=1+M A(r,\alpha)$, we have
\begin{equation}
    \left(\frac{\partial^2 f}{\partial M\,\partial r}\right)_{\alpha}
    =
    A'(r,\alpha)
    =
    \frac{1}{M_h}
    \left(\frac{\partial f}{\partial r}\right)_{M,\alpha}
    =
    \frac{\mathcal F}{M_h}.
\end{equation}
Using Eq.~\eqref{Mprime}, we finally obtain
\begin{equation}
    \dot{\mathcal F}(r_h,\alpha)
    =
    \left(\frac{\partial^2 f}{\partial r^2}\right)_{M,\alpha}
    \bigg|_{r=r_h,\,M=M_h}
    +
    \mathcal F^2(r_h,\alpha).
    \label{Fdot_def}
\end{equation}
This extra term $\mathcal F^2$ is a consequence of the fact that the
mass parameter changes with $r_h$ along the horizon branch.

Explicitly,
\begin{align}
\dot{\mathcal F}(r_h,\alpha)
&=
\mathcal F^2(r_h,\alpha)
+
M_h e^{-r_h/\alpha}\notag\\
&\times
\left(
\frac{4}{r_h^3}
+\frac{4}{\alpha r_h^2}
+\frac{2}{\alpha^2 r_h}
+\frac{r_h}{\alpha^4}
\right)
-\frac{4M_h}{r_h^3}.
\label{Fdot_explicit}
\end{align}

Equations~\eqref{aa2a} and~\eqref{Fdot_explicit} give the specific heat
capacity in a form that makes clear which derivative is being used. In
the Schwarzschild limit, $\alpha\to0^+$, one has
\begin{equation}
    \mathcal F(r_h)=\frac{1}{r_h},
    \qquad
    \dot{\mathcal F}(r_h)=-\frac{1}{r_h^2},
\end{equation}
and therefore Eq.~\eqref{aa2a} reduces to
\begin{equation}
    C_{\rm Sch}
    =
    2\pi r_h
    \frac{1/r_h}{-1/r_h^2}
    =
    -2\pi r_h^2,
\end{equation}
as expected.

\begin{figure}[htbp]
\centering
\includegraphics[width=\columnwidth]{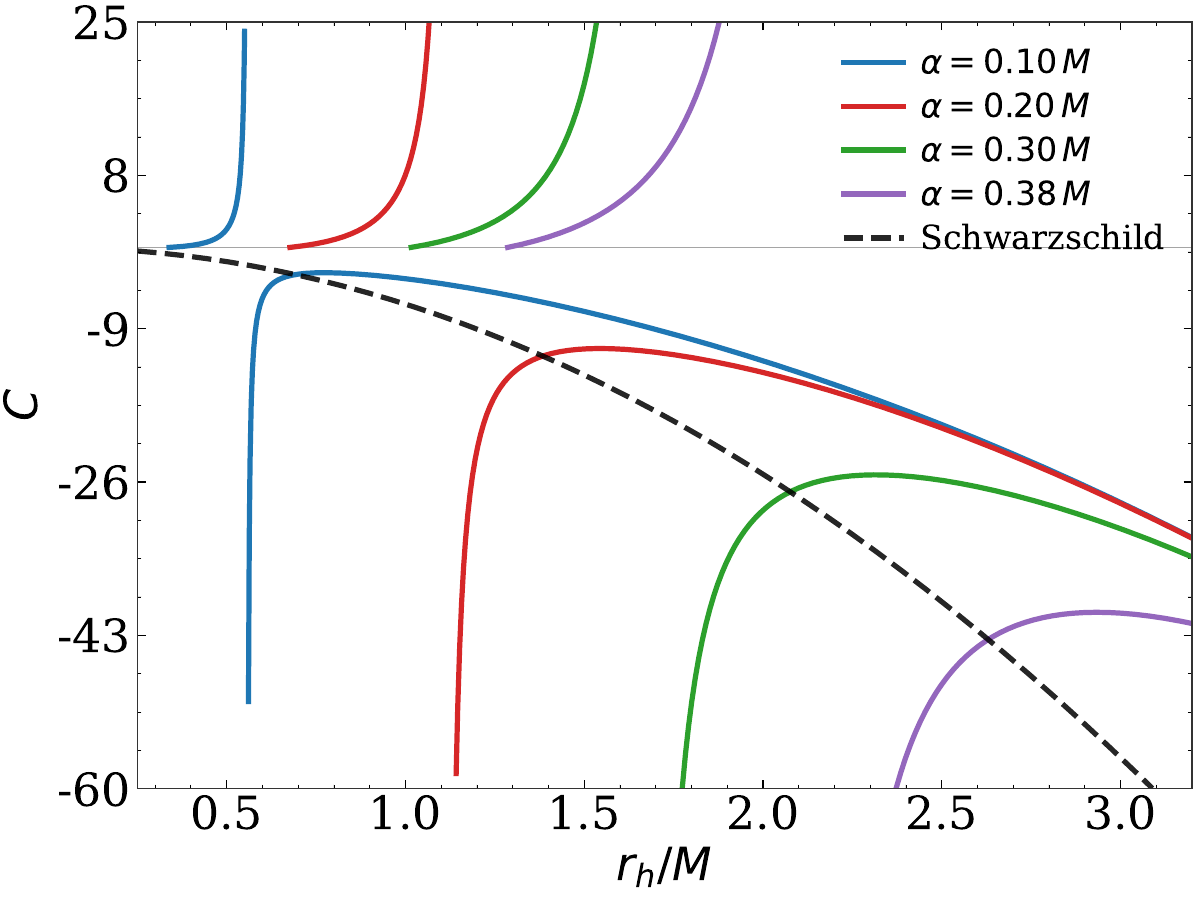}
\caption{Specific heat capacity $C$ as a function of $r_h$ for
$\alpha/M=0.10,\,0.20,\,0.30,$ and $0.38$. The dashed black curve shows
the Schwarzschild result $C_{\rm Sch}=-2\pi r_h^2$, which is everywhere
negative. The regular black-hole solutions exhibit a positive-$C$
(stable) branch at small $r_h$, followed by a Davies transition where
$C$ diverges, and the branch becomes unstable. As $\alpha$ increases, the
stable region extends to larger horizon radii.}
\label{fig:specific_heat}
\end{figure}

The sign of the specific heat determines thermodynamic stability:
$C>0$ corresponds to a stable configuration, while $C<0$ indicates
instability. Divergences of $C$, equivalently zeros of
$\dot{\mathcal F}(r_h,\alpha)$, signal phase transitions of Davies type
\cite{HawkingPage1983,Davies1978}. Figure~\ref{fig:specific_heat}
shows $C$ as a function of $r_h$ for representative values of $\alpha$.
In contrast to the Schwarzschild case, where $C_{\rm Sch}=-2\pi r_h^2$ is always negative, the dark matter regular black hole exhibits a positive-$C$ region for sufficiently small $r_h$, signaling a thermodynamically stable phase. The boundary between the stable and unstable regions defines a phase transition, and the extent of the stable phase broadens as $\alpha$ increases.

\section{Modified First Law of Thermodynamics and Smarr Relation}

As stated earlier, we considered that the black hole thermodynamic system obeys the Bekenstein-Hawking entropy formula given by \cite{Bekenstein1973,Hawking1975} \(S=\pi r_h^2\). Assuming that the scale parameter $\alpha$ behaves as a thermodynamic intensive variable. The black hole mass $M$ in (\ref{mass}) is a function of $S$ and $\alpha$, that is, $M=M(S, \alpha)$. Therefore, the modified first law of thermodynamics \cite{BardeenCarterHawking1973} can be expressed as,
\begin{equation}
    dM=T dS+\Phi d\alpha,\label{ww2}
\end{equation}
where the thermodynamic temperature $T$ is given by
\begin{align}
    T=\left(\frac{\partial M}{\partial S}\right)_{\alpha}=\frac{1}{2 \pi r_h} \left(\frac{\partial M}{\partial r_h}\right)_{\alpha}=T_H\label{ww3}
\end{align}
which is equal to the geometric temperature (called the Hawking temperature).

And the conjugate potential $\Phi$ associated with $\alpha$ as
\begin{equation}
    \Phi=\left(\frac{\partial M}{\partial \alpha}\right)_{S}= M^2 e^{-r_h/\alpha} \frac{r_h^2}{\alpha^4}.\label{ww4}
\end{equation}

The Smarr relation arises from scaling symmetry, which can be derived either using Euler’s theorem for homogeneous functions or Komar's integral methods. Since the mass is homogeneous of degree one under the simultaneous scaling $(r_h, \alpha) \to (\lambda r_h, \lambda \alpha)$, the system effectively retains scale invariance in this generalized sense. Consequently, the Smarr relation \cite{Smarr1973} takes the form
\begin{equation}
    M = 2TS + \alpha\,\Phi,
    \label{ww5}
\end{equation}
which holds for the present case.

Noted that the first law follows from the horizon condition and remains valid, while the Smarr relation is modified due to the presence of the scale $\alpha$ appearing in the exponential term. From the above analysis, we have observed that the thermodynamic system of the considered regular black hole satisfies the first law of thermodynamics, analogous to the Bardeen/Hayward/Dymnikova models, and also holds the Smarr relation.

\section{Sparsity of Radiation}\label{sec:3}

Although Hawking radiation exhibits a thermal spectrum, the emission
process is not continuous in time. Instead, it proceeds through the
emission of discrete, well-separated quanta, implying that the Hawking
flux is intrinsically sparse \cite{Visser2017MGM,Gray2016}. A useful
way to quantify this sparsity is by comparing the typical thermal
wavelength of the emitted particles with the effective emission area of
the black hole \cite{Visser2017MGM}.

The sparsity parameter is defined as
\begin{equation}
    \eta=\frac{\mathcal{C}}{\tilde{g}}\,\frac{\lambda^2_{\rm th}}
    {A_{\rm eff}},\label{bb1}
\end{equation}
where $\mathcal{C}$ is a positive numerical constant, $\tilde{g}$ is the
number of spin states of the emitted quanta, and
\begin{equation}
    \lambda_{\rm th}=\frac{2 \pi}{T_H},\qquad
    A_{\rm eff}=\frac{27}{4}\,A_{\rm BH}(r_h)
              = 27\pi r_h^2.\label{bb2}
\end{equation}
Here $A_{\rm BH}(r_h)=4\pi r_h^2$ is the horizon area. Using
$T_H=\mathcal F(r_h,\alpha)/(4\pi)$ from
Eqs.~\eqref{F_def} and~\eqref{F_explicit}, and setting
$\mathcal{C}/\tilde{g}=1$ for a single bosonic species, the sparsity
parameter becomes
\begin{equation}
\eta=
\frac{64 \pi^3}{27}\,
\frac{1}{r_h^2\,\mathcal F^2(r_h,\alpha)}.
\label{bb3}
\end{equation}
In the Schwarzschild limit, where $\alpha\to0^+$, one has
$\mathcal F(r_h)=1/r_h$, and therefore
\begin{equation}
    \eta_{\rm Sch}=\frac{64\pi^3}{27}.
\end{equation}

\begin{figure}[htbp]
\centering
\includegraphics[width=\columnwidth]{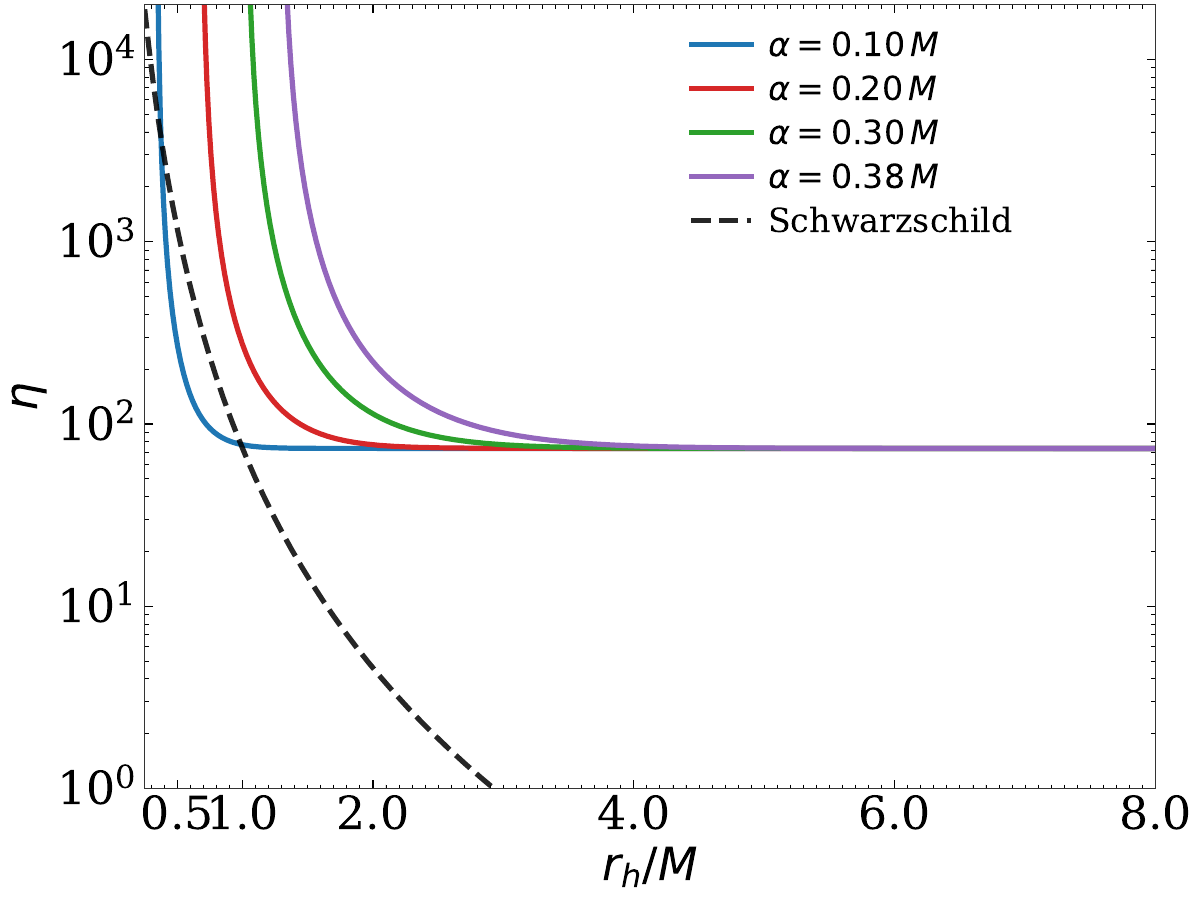}
\caption{Sparsity parameter $\eta$ (log scale) as a function of $r_h$
for $\alpha/M=0.10,\,0.20,\,0.30,$ and $0.38$. The dashed black curve
shows the Schwarzschild result $\eta_{\rm Sch}=64\pi^3/27$. A larger
$\eta$ indicates a more intermittent Hawking flux. The halo correction
enhances the sparsity relative to Schwarzschild, and the enhancement
becomes stronger as $\alpha$ increases.}
\label{fig:sparsity}
\end{figure}

Figure~\ref{fig:sparsity} displays $\eta$ on a logarithmic scale as a
function of $r_h$. For all values of $\alpha$ the sparsity parameter
satisfies $\eta\gg 1$, confirming that Hawking radiation is far from
a continuous blackbody stream. Furthermore, for a fixed horizon radius
the sparsity is larger for larger $\alpha$, i.e., a stronger halo
correction makes the Hawking flux more intermittent. This can be
understood from Eq.~\eqref{bb3}: a reduced temperature, equivalently a
smaller $\mathcal F(r_h,\alpha)$, increases $\eta$ for fixed $r_h$.
As $\alpha\to0^+$, all curves converge to the Schwarzschild sparsity.

\begin{figure}[htbp]
\centering
\includegraphics[width=\columnwidth]{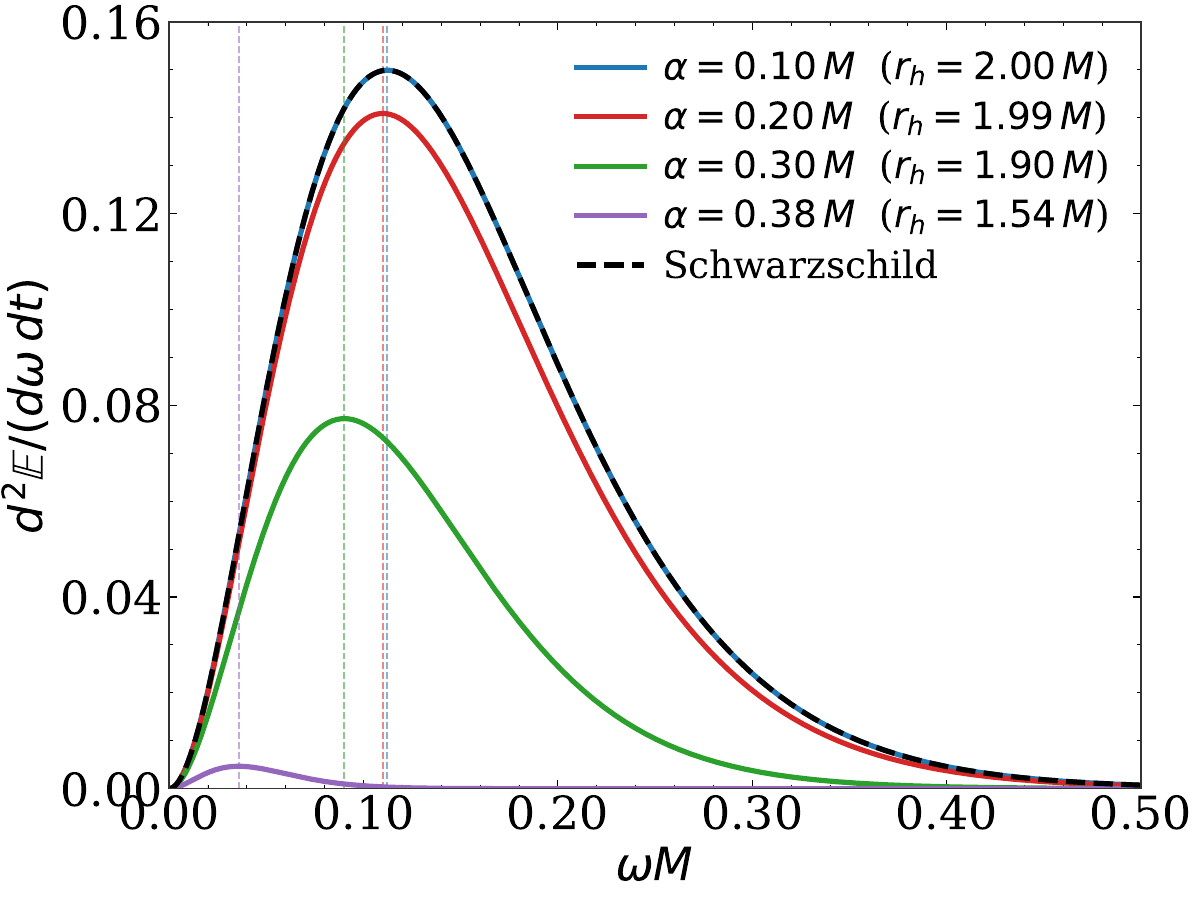}
\caption{Spectral energy emission rate $d^2\mathbb{E}/(d\omega\,dt)$
as a function of emitted frequency $\omega$ (units of $M^{-1}$) for $M=1$ and $\alpha/M=0.10,\,0.20,\,0.30,$ and $0.38$. The dashed black curve is the Schwarzschild result, while the legend also indicates the corresponding horizon radius for each coloured branch. Larger $\alpha$ suppresses and red-shifts the emission spectrum. Vertical dashed lines mark the peak frequency of each curve.}
\label{fig:emission}
\end{figure}

\section{Energy Emission Rate}\label{sec:4}

Before deriving the energy emission rate, we first determine the black hole's shadow radius. For an asymptotically flat spacetime, the shadow radius for a distant static observer is
\cite{Perlick2022PhysRep,Cunha2018Shadows}
\begin{equation}
    R_{\rm sh}=\frac{r_{p}}{\sqrt{f(r_p)}},\label{cc1}
\end{equation}
where $r_p$ is the photon-sphere radius, determined by the circular
null-geodesic condition
\begin{equation}
    2\,f(r_p)-r_p\,f'(r_p)=0.\label{cc2}
\end{equation}
For Schwarzschild, one recovers $r_p=3M$ and
$R_{\rm sh}=3\sqrt{3}\,M$.

In the high-frequency (geometric-optics) regime, the absorption
cross-section oscillates around a limiting constant value
\cite{Mashhoon1973,Sanchez1978,Decanini2011,Wei2013}. Since the
capture of high-energy quanta is governed by null geodesics, the limiting absorption cross-section is estimated as \cite{Wei2013}
\begin{equation}
\sigma_{\rm lim}\approx \pi R_{\rm sh}^2.\label{cc3}
\end{equation}
Within this approximation, the spectral energy emission rate is
\cite{Wei2013,Sanchez1978,Decanini2011}
\begin{equation}
\frac{d^2\mathbb{E}}{d\omega\,dt}
  =
\frac{2\pi^2\,\sigma_{\rm lim}}{e^{\omega/T_H}-1}\,\omega^3,
  \label{cc4}
\end{equation}
where $\omega$ denotes the emitted frequency. Substituting
Eq.~\eqref{cc3} and using
$T_H=\mathcal F(r_h,\alpha)/(4\pi)$, we obtain the compact expression
\begin{equation}
\frac{d^2\mathbb{E}}{d\omega\,dt}
  =
\frac{2\pi^3\,R_{\rm sh}^2\,\omega^3}
{e^{4\pi\omega/\mathcal F(r_h,\alpha)}-1}.
  \label{cc5}
\end{equation}

Figure~\ref{fig:emission} shows the spectral emission rate as a function
of frequency $\omega$ for $M=1$ and representative values of $\alpha$. As
$\alpha$ increases: (i) the peak emission rate diminishes, (ii) the
peak shifts to lower frequencies, reflecting the lower temperature, and
(iii) the overall emitted power is suppressed, consistent with a cooler
and a less luminous black hole. In the limit $\alpha\to0^+$, the
Schwarzschild curve, with $r_h=2M$, $T_H=1/(8\pi M)$, and
$R_{\rm sh}=3\sqrt{3}\,M$, is recovered.

\section{Conclusion}\label{sec:5}

In this work, we have investigated the thermodynamic and radiative properties of a singularity-free black hole sourced by a dark matter halo described by the Einasto density profile. Starting from the exact regular solution of Konoplya and Zhidenko \cite{Konoplya2026}, we derived analytical expressions for the Hawking temperature, the specific heat capacity, the sparsity of the Hawking flux, and the
spectral energy emission rate, and analyzed their dependence on the characteristic Einasto scale parameter $\alpha$.

Our principal findings can be summarised as follows.

\emph{Hawking temperature.} The presence of the dark matter halo suppresses the Hawking temperature below the Schwarzschild value for the same horizon radius, with the suppression increasing as $\alpha$ increases. This behavior reflects the additional negative contribution from the halo to the derivative of the lapse function. In the limit $\alpha\to 0$ the standard Schwarzschild temperature
$T_{\rm Sch}=1/(8\pi M)$ is recovered.

\emph{Specific heat and thermodynamic stability.} Whereas the Schwarzschild black hole is thermodynamically unstable ($C<0$) for all horizon radii, the dark matter regular black hole exhibits a region of positive specific heat ($C>0$) for sufficiently small $r_h$. The boundary between the stable and unstable phases constitutes a Davies phase transition, signaled by the divergence of $C$. The extent of the thermodynamically stable region grows as $\alpha$ increases,
suggesting that a more concentrated dark matter distribution enhances the stability of small black holes. This is one of the most physically distinctive features of the dark-matter regular black hole compared with its classical counterpart.

\emph{Sparsity.} The sparsity parameter $\eta$ satisfies $\eta\gg 1$ for all configurations studied, confirming the highly intermittent, non-continuous character of Hawking radiation in this spacetime. For fixed $r_h$, the sparsity is larger for larger $\alpha$, meaning that a stronger halo correction makes the Hawking flux sparser. This enhancement follows directly from the suppression of the temperature: a cooler black hole emits rarer quanta.

\emph{Energy emission rate.} The dark matter halo suppresses and redshifts the spectral energy emission rate: both the peak value and the peak frequency decrease as $\alpha$ increases. This implies that black holes embedded in denser Einasto halos are dimmer and emit at lower characteristic frequencies than their Schwarzschild counterparts.

All results reduce smoothly to the Schwarzschild expressions in the limit $\alpha\to 0$, providing a non-trivial consistency check of the entire analysis.

From a broader perspective, the present results demonstrate that the dark matter environment surrounding a black hole leaves quantitative imprints not only on its gravitational-wave spectrum or shadow size, but also on its thermodynamic and radiative properties. These imprints could, in principle, be constrained by future observations of black hole luminosities and spectra, or by measurements of the gravitational-wave background from black hole evaporation scenarios. Extensions of this analysis to rotating (Kerr-like) regular black holes embedded in dark
matter halos, and to quantum-corrected scenarios, are left for future
work.

\scriptsize
\section*{Acknowledgments}

F.A. acknowledges the Inter University Centre for Astronomy and Astrophysics (IUCAA), Pune, India for granting visiting associateship.  E. O. Silva acknowledges the support from Conselho Nacional de Desenvolvimento Cient\'{i}fico e Tecnol\'{o}gico (CNPq) (grants 306308/2022-3), Funda\c c\~ao de Amparo \`{a} Pesquisa e ao Desenvolvimento Cient\'{i}fico e Tecnol\'{o}gico do Maranh\~ao (FAPEMA) (grants UNIVERSAL-06395/22), and Coordena\c c\~ao de Aperfei\c coamento de Pessoal de N\'{i}vel Superior (CAPES) - Brazil (Code 001).


%

\end{document}